# ARCHAEOASTRONOMY
# IN THE KHMER HEARTLAND

Giulio Magli
School of Architecture, Urban Planning and Construction Engineering,
Politecnico di Milano, Italy
Giulio.Magli@polimi.it

*The heartland of the Khmer empire is literally crowded by magnificent monuments built in the course of many centuries. These monuments include the world-famous "state-temples", such as Angkor Wat, but also many other temples and huge water reservoirs. Using Google Earth data as well as GIS data and reconstructing the ancient sky with Stellarium, we investigate here on the relationships of astronomy with orientation and topography in a systematic fashion, following the methods of modern Archaeoastronomy and strictly keeping at a bay vague and/or esoteric proposals put forward by many authors in the past. As a result, a very clear pattern of cardinal orientation and alignment arises, connected with the temple's symbolism and the management of power by the Khmer kings. As a bonus, the comparison with the Angkor monuments allows to put forward a explanation for the anomalous orientation of the unique two "peripheral" state temples of Cambodia.*

## 1. Introduction

The Khmer empire flourished between the 8th and the 14th century AD. The heartland of the empire was in the vast Cambodian lowlands, where the kings adopted monumental temple architecture as a means for the explicit representation of their power; as a consequence, a series of masterpieces – and especially the so called "state temples", like Angkor Wat - were constructed (Jacques and Lafond 2007). Geographically, these buildings concentrated in the surroundings of today's' Siem Reap, first in the area of Roulos, while later the kings moved to Angkor, some 15 kilometres to the north. There are, however, two exceptions: Koh Ker, located in northern Cambodia 85 kilometres north-east of Angkor, and Preah Khan of Kompong Svay, 100 Kms to the east.[1]
The Khmer "state" temples are vast rectangular enclosures enclosing a central unit and several auxiliary buildings and shrines. The aims of such architectural ensembles, whose project in many cases included also the construction huge *barays* (water reservoirs), were quite complex, as they functioned as royal residence and main centre of cult attesting to the beliefs and religiosity of the king. A further, funerary function for the after death of the king, although likely, has never been proved. Up to a few years ago they were even conceived as "concentrated state towns" but recent research and mapping on large scale has shown the complexity of the urbanization of the whole Angkor area, putting aside the idea of the state-temples as “capital cities” (Fletcher and Evans 2015, Stark et al. 2015).

[1] The functioning of Preah Khan of Kompong Svay as state-temple is still debated; see section 6.

Another important aspect put in evidence by recent research is the high degree of sophistication of the hydraulic system, which led to an impressive modification of the natural environment. It can be thought of as organized into 3 grand areas, with the major barays acting as central collectors and flow management systems towards the south. The barays thus had both a practical and ritualistic function, in being explicitly associated with the state temples and embellished with the Mebons, the island temples built inside them (Fletcher et al 2015).

As far as the interpretation of each temple is concerned, the complexity of the relationship between the two Khmer religions - Buddhism and Hinduism – must be taken into account. This relation was sometimes exclusive and other times sincretist, in dependence of the attitudes of the ruling king. Documented historical phases of Buddhism/Hinduism explicit conflict and consequent defacing of temples' images also exist. In any case, construction of the temples was clearly considered as mandatory to attest to the greatness and in some sense to the divinity of the king himself. The temples thus reflected concepts related to the foundation of power and to the cosmic order; as a consequence, it comes as no surprise that a complex religious symbolism is self-evident in all these buildings. Inspiration certainly was in Indian sacred architecture (see e.g. Malville and Gujral 2000, Kak 1999, 2001) and in particular, the characteristic layout of Angkor Wat and of many other temples - a “pyramid mountain” surrounded by a moat—is usually considered to correspond with the cosmology of Mount Meru and the surrounding Sea of Milk from which ambrosia was churned by the gods and demons.

The religious symbolism associated with cosmic order is reflected in the orientations of ancient buildings worldwide (see e.g. Magli 2015), and Angkor is not an exception. It is indeed well known that almost all the temples, enclosures and barays are oriented cardinally. However, although this notion is widespread (see e.g. Petrotchenko 2012), as far as the present author is aware no attempt has been made to analyse it quantitatively, using the methods of modern Archaeaostronomy (see e.g. Ruggles 2015, Magli 2015). Further – and curiously – the same pattern is *not* respected at the two complexes located out of the heartland, so that their orientations can be defined as “anomalous”. Also this fact has been noticed in the literature, but never explained. Yet another point which has been left unexamined is the reported existence of alignments between different temples.

In spite of (or perhaps due to) this lack of academic archaeoastronomical studies, pseudo-archaology literature on Angkor easily finds his way out on international media. The same holds for the “astronomical numerology” of Angkor Wat which has been made famous by a controversial book (Mannika 1996). The present author holds many reserves on this issue, but discussing it would be out of the scope of the present paper.

To study the Archaeoastronomy of the Angkor temples, a complete database of orientations at Angkor has been constructed here using satellite imagery, and the sky over Angkor has been reconstructed using the potentialities of the software Stellarium. The database (Table 1) is presented in a chronological fashion (the reader is, however, advised that not all the attributions of the temples are firmly established in the archaeological literature). This allows to investigate not only orientations but also the likeness of intended alignments between subsequent buildings. By comparison with the Angkor results, the anomalous orientations of the peripheral temples can also be interpreted as representing two different breaks trough the traditional pattern of orientation.

## 2. The orientations of the Angkor monuments

The author has taken in the past sample data of Angkor temples with a precision magnetic compass. However, in order to present a complete analysis based on a homogeneous, reliable and complete set of data, the azimuths used in this paper (reported in Table 1) have all been obtained with the compass tool of Google Earth Pro. The accuracy of Google Earth in areas covered by high-resolution images is usually very good (Potere 2008), and in particular - as the author has verified personally comparing satellite data with directly acquired theodolite measures in many different countries – the error in azimuth does not exceed ±½°. The reliability of this estimate is particularly solid in the present case because all the data have been subjected to a double-blind control. In fact the author was kindly allowed to consult the GIS database currently being developed by the *Greater Angkor Project,* and in all cases the azimuth measures furnished by this database were very close to those of Google Earth; the two "virtual campaigns" have of course been done independently, that is, without mutually adjusting the way of taking measures. As far as the horizon is concerned, it is flat for almost all monuments (the case of Angkor Wat will be treated separately) and, as we shall see, the monuments by themselves actually work as artificial horizons.

The following monuments have been listed and measured:

1) State temples

2) Barays. I opted to consider the Barays as monuments on their own, since their size and meticulous orientation are difficult to reconcile with purely functional aims. Actually the results of the paper support the view that they were an essential part of the building programme associated with the legitimization of kingship.

3) Island temples, constructed in many cases in already existing barays, typically by successors

4) The most important secondary temples, either of royal or private construction.

The main entrance of all the temples, excluding Angkor Wat and the secondary temple of Wat Athvear constructed by the same king, is to the east. For each temple, a convention similar to that commonly used for Greek temples has thus been followed, so that the azimuth from inside looking out is given. The results (Table 1, reported schematically in Fig. 1) show that there is a obvious pattern of orientation towards the true east exhibited by all the data (since the rectangular enclosures are sometimes not perfectly squared, in the table both azimuths of the east-west and of the north-south sides are reported, but the analysis is based on the azimuths of the east-west sides). *All* of the 31 monuments considered lie within an interval as small as five degrees, between 85° and 90°. It is obvious that there is no need of statistics to conclude that the orientation was intentional. A first point is, therefore, fully confirmed: these monuments were connected with ideas of cosmic order in architecture, which imposed orientation to the cardinal points. However, the data tell us more than that. Indeed, 19 monuments out of 31 concentrate between 89 and 90 degrees, and *all* those temples not perfectly oriented to due east exhibit a slight deviation exclusively to the *north* of east; not even one exhibits a deviation, whatever small, to the south of east. Are these small deviations only due to errors committed by the builders in the measurement process, or instead they are intended? It is clear from the monuments themselves that the Khmer architects were extremely precise. The method they used to find the cardinal directions was probably based on the sun, both because of religious reasons as well as practical reasons. Indeed finding north using the stars requires either the observation of the directions of rising and setting of a bright star on a flat, levelled horizon (a thing quite difficult to realize in the humid environment of Angkor) or the observation of the motion of a circumpolar star (again, very difficult due to the very low height of the north celestial pole). Further of course, due to precession, no "pole star" was available in Khmer times. On the other end, the traditional method of finding cardinal directions by bisecting the shadows of a gnomon on a circle – the "Indian circle" - was certainly well known to the Khmers. Using this method, a scrupulous astronomer can easily reach an accuracy of the order of ½°, if not better. We are thus led to think that even the slight deviations observed in some temples are deliberate. Further, if the observed deviations were originated by errors of measure, then either a method which leads *only* to north-of-east errors was devised (a thing which looks unlikely) or the results should distribute randomly on both sides of the expected value of 90°.
The key to this riddle can be found studying the orientation of Angkor Wat, as we shall now discuss. The azimuth of the temple is 270.5°. The likely reason is that the temple was originally dedicated to Vishnu, a God tightly connected to the west, as can be seen, for instance, in the upper terrace of the Bayon, whose western shrine is devoted to him. The orientation of Angkor of course implies that a person entering the enclosure from the west gate is looking along the direction of azimuth 90.5°. This slight deviation has an interesting consequence, which is already very well known (Stencel et al 1976). Looking from the west gate towards the temple at dawn at the equinoxes, the sun is seen to rise just above the central tower, "crowning" it almost vertically. The reason is that at the latitude of Angkor the trajectory of the sun is very steep, and therefore a small increase in azimuth leads to a strong increase in height; the "horizon height" of the central tower of Angkor Vat from the western entrance is ~5° and the centre of the sun reaches such an altitude at an azimuth of 90° 40' (Fig. 2).[2]
So far so good for the Angkor Wat orientation. It is now obvious, however, that if a similar phenomenon has to be observed in a temple whose main access is to the east, observation will occur at sunset, and the azimuth of the temple must be slightly misaligned *to the north of east*. In fact in this way the direction of the observer looking along the temple axis will point slightly *to the south of west*, where the equinoctial sun will be seen to disappear just above the temple.
This is the likely explanation for the temples oriented slightly to the north of east, and so in particular for the state temples Bakong, Phnom Bakheng and Bayon, whose azimuth ("exactly" as that of Angkor Wat, but in the opposite direction) is 89.5°. The phenomenon of the sun disappearing vertically behind the temple at the equinoxes in this latter case can be verified using Google Earth 3D visualization. The dimension of the sun in Google Earth simulations are relatively big, but in spite of this the effect is unmistakable, as shown in Fig. 3.
What about the meaning of such spectacular hierophanies? For Angkor Wat, Stencel et al. (1976) proposed a rather complex calendrical function, trying to frame the phenomenon into a series of supposed astronomical functions of the monument which, when viewed from different, suitably chosen points of the esplanade should furnish, for instance, the extrema of the motion of the Moon at the horizon. This idea is however clealry biased by a strong selection effect of the observation points and in any case there is no evidence – and indeed, it is unlikely - that the Khmer monuments were used for precise astronomical observations. The key is instead symbolic: the beautiful hierophany of the sun suspended just above the mountain-temples at the equinoxes was very probably intended as a materialization of the connection of the temple itself with the heavens, since it realizes a match between the cardinal directions on earth and the zenith. To back up this interpretation, a well known general connection of the temple-mountain architecture with the axis orthogonal to the earth's surface - the zenith-nadir axis to which respectively the temple and its image reflected in the waters in front of it allude explicitly - can be summed up with the recent discovery that the zenith passages were probably also made visible *inside* the temples (Barnhart and Powell 2015). In fact, although the simplest way to observe the zenith passages is to look for the days in which the shadow of a post vanishes at noon, another efficient way is to use a straight vertical tube leading from the open sky into a dark chamber. If the tube is sufficiently long and narrow, the identification of the zenith passage will be accurate and – most of all - the effect inside the room will be spectacular. This method, which was devised for instance in Mesoamerica in Columbian times, was very likely in use in the Angkor temples. Today, their roofs is open, but the capstones are missing. Many of such stones are however present in the rubble near the temples, and all have a hollow tube running down their axes. The holes allowed rain to hit the holly stone *lingas* located at the centre of many chambers, but also allowed the sun passing overhead to light the same stones in spectacular hierophanies occurring twice a year (zenith passages at Angkor occur on April 26 and August 17).
The temples of the Angkor heartland were thus anchored with the cycle of the sun in two ways: the orientation, related to the equinoxes, and the vertical openings of their chambers, related to zenith passage. Yet another way to connect a temple with the zenith passages is, of course, that of orienting the building to the sun rising or setting in these days, as occurs, for instance, for the world famous post-classic Maya pyramid of Chichen Itza, Yucatan. The azimuth of the rising sun on the zenith passages at Angkor is 76°, so apparently no Angkor temple was oriented in this way. Interestingly enough, however, the cardinal orientation appears to be a pattern, a rule, for the sacred space of Angkor only, since – as we shall see in section 6 – a state temple oriented to the sun rising on the days of the zenith passages actually exists out of the heartland.
Finally, a few temples remain for which the misalignment is too elevated to be advocated to visual effects considerations. In particular the temple exhibiting the worse misalignment with respect to due east, Banteay Kdei, was probably constructed much after the construction of the baray which lies in front of it and has the same alignment. Thus, probably the original rough orientation of the baray influenced that of the later temple for aesthetic reasons.

[2] Stiff astronomical data in this paper are taken from Stellarium, while declinations are calculated with the program Get-Dec kindly provided by C. Ruggles, which takes into account refraction and parallax.

## 3. Astronomical alignments between monuments

The existence of scores of alignments between different monuments of Angkor was noticed many years ago by Paris (1941), who offered however no explanation for them. He divided his finds in cardinal (north-south or east-west) alignments, sosticial alignments, and "non-oriented" (meaning at least 3 points aligned but not astronomically) alignments. In total, he proposed 28 cardinal alignments, 26 sosticial alignments, and 11 other alignments. In the present paper we are interested in astronomical alignments only, and therefore I shall consider further only cardinal and sosticial relationships. [3]
I have subjected all the proposed alignments to an accurate check. The results are the following:

1) A few are impossible to verify, as they refer to unnamed buildings, or ruins which are not recognizable.
2) A few others are not verified within the error allowed for here, namely ½°.
3) All the remaining ones are "correct"; in other words, within the accuracy adopted in this paper they are indeed verified.

Among the alignments which are technically verified, the following possibilities may occur:

1) Alignments occurring by pure chance.
These alignments arise due to a selection effect. For example, a side of a temple complex is aligned with a corner of another complex and with the opposite side of yet another one, a connection which is far more easy to occur by chance than – say - to find out that the tops of 3 temples are all aligned on the same meridian. In particular, special attention must be exerted when the temples are too far and do not allow a direct view. Indeed, due to the Earth roundness, inter-visibility between sites (provided that the view is unobstructed) is severely limited. A good estimate is the following: the visibility of an object which is h meters high equals the square root of 13 h expressed in kilometres, so that, for instance, a person 2 meters tall sees on a flat horizon at about 5 Kms distance. The summits of existing temples and/or provisional wooden structures could have been used to trace more long alignments, because when the object sighted is in itself high, the heights add each other and therefore the horizon distance increases. However, also in this case, lines longer than (say) 11-12 Kms (corresponding to observation points located at heights ~10 meters) must be regarded as extremely suspicious.

[3] Paris also proposed geometric relationships between temples (like e.g. temples standing at the 3 vertexes of equilateral triangle or the 4 vertexes of a trapezoid), which will not be investigated here.

2) Alignments occurring for technical reasons.

The technical problem of tracing cardinally oriented lines is not an easy task; it does not suffice to determine the cardinal directions with a suitable accuracy, it is also needed to keep it with the same accuracy during the construction of – say – a tremendous work of engineering such as the West Baray, which is 7.8 Kms long. This has the consequence that already surveyed lines can be useful for a purely practical viewpoint; for example, the top of a temple can be used as a survey point when tracing a new project. This procedure generates "true" alignments whose explanation is, however, purely functional.

3) Alignments occurring for symbolic reasons.

For example, a certain project was connected to another trough an astronomical event, or a certain king wanted to create a visual connection with the monument of a previous king. This kind of alignments are, of course, the unique of interest for Archaeoastronomy.

## 3.1 Cardinal alignments

After a careful scrutiny, only a certain number of the cardinal alignments proposed by Paris survived to the "selection effect" test. Further, claiming intentionality according to Point 3 above is possible only if the chronology of the monuments in question is perfectly clear. For this reason, in what follows, only alignments including a well attested chronology are reported, although I do not exclude the possible existence of a few others involving other buildings, whose chronology is not yet clear.

Our starting point is the first state temple built in Angkor-Roulos, the Bakong, built by Indravarman I. The temple was built along with a huge baray, the so-called Indrataka, which was located in such a way that the central axis lies on the same meridian (azimuth 180°) of the Bakong, located 1.8 Kms to the south of the baray's centre. The successor of Indravarman I, Yasovarman I, built *inside* the already existing Baray a island temple, the Lolei, which was placed along the very same meridian (Fig. 4). In this way, a – clearly intentional and symbolic – alignment was realized.

The same king also "delimited" the sacred space of the future Angkor by building two temples (Phnom Krom and Phnom Bok) on the two small hills which overlook the area for the south and from the east respectively.

Around the beginning of the $10^{th}$ century AD Yasovarman I probably initiated also the construction of a even huger baray, the East Baray, which was – about 50 years later - used by Rajendravarman II create a symbolic configuration extremely similar to that of the Lolei-Bakong. Indeed he built his own state temple, Pre Rup, on the same meridian of the centre of the East Baray, and about 1.3 Kms to the south. Further, he added an island temple (the East Mebon) inside the already existing baray. The position chosen for the East Mebon is "on the same meridian" of Pre Rup. The reason for the quotation marks is that the azimuth of this alignment is slightly out of centre (178°), but this is coherent with the orientation of Pre-Rup which indeed is 88°. In other words, it is clear that East Mebon was intentionally placed along the direction orthogonal to the north side of Pre Rup. Interestingly, a very long line (about 7 Kms) almost lying on the parallel (azimuth 269°) further connects East Mebon with Phimeneakeas, whose original dating is unsure but which functioned as the state temple of the successor Suryavarman I. The Victory Gate of Angkor Thom lies along the very same direction and its placement was almost certainly chosen for the very same reason (Fig. 5).

Finally, a interesting combination of alignments also repeats for the hugest of the barays, the West Baray. Here the north side is on the same parallel of the state temple of its builder, the Baphuon, which lies at a distance of 2.2 Kms from the corner, to the east. The prolongation of the south side to the east also lies on the same parallel (azimuth 89.5°) of a temple, the pre-existing Bat Chum, at some 7.6 Kms from the south-east corner. Finally at the centre of the West Baray the island temple West Mebon was constructed by Udayadityavarman II (it is not completely clear if the project of the baray was already initiated by his predecessor Suryavarman I). The west Mebon is connected with Pre Rup, since a "parallel" line (azimuth 89.5°) connects the two monuments (later, also Ta Prohm will be built on this line). This line is very long (9.6 Kms) but Pre Rup is 12 meters high. Clearly however, since the Mebon is at the centre of the baray, it is the baray itself to have been planned taking into account 3 reference points corresponding to the parallels to the sides and the mid line (Fig. 5).

The reasons for the construction of the barays (whether they were purely functional, or purely ritualistic, or both) is a complex issue of Khmer archaeology as a whole and certainly cannot be addressed by the present author. Only, I would like to put in evidence that the existence of the above mentioned, certainly intended alignments reinforces the presence of a ritual function, as there were the Khmer kings by themselves to put these alignments in clear evidence by the construction of the island temples. Building such monuments was certainly not an easy task indeed.[4] It is difficult to believe that the islands were rapidly assembled during the dry season (and further, it is unclear to what extent the barays were dried during this season) and it is even more difficult to believe that they were built within the waters, so the barays must have been dried voluntarily (by closing the inlet moats) for a suitable time. Their construction therefore certainly expressed a direct will by the kings, and was due to symbolic reasons. The existence of the above described topographical relationships of the barays with the state temple of their builders allowed to construct the Mebons *in accordance with existing lines of sight oriented cardinally*. The idea of constructing the Island temples appears therefore to implement in a quite spectacular and sophisticated way a dynastic *continuity* which the kings wanted to make explicit (Fig. 4,5). Creating inter-connecting, visual relationships with monuments built by predecessors is a means of stating the continuity of power and the "divine" rights of ruleship, according to a mechanism of development of the topography of the sacred space which is similar to those developed in completely unrelated places and times (see e.g. Aveni and Hartung 1988 for the Mayas, Magli 2010 for Old Kingdom Egypt). Clearly, to this aim, the existing alignments between barays and temples were easily operational.

---

[4] A Mebon (Neak Pean) was also built inside the baray of the Preah Khan temple, probably within the same project.

From the point of view of "dynastic" cardinal alignments, there exists two other cases which convincingly look as being non-casual. These are:

- A very precise (azimuth 89.5°) east-west line which connects the top of the Bayon with that of the pre-existing temple Banteay Samrè. The axis runs along the "Gate of the deaths" of Angkor Thom. It is very long (about 10.8 Kms, as it crosses the whole of the East Baray) but the top of Banteay Samre' could be used as a survey point (Fig. 4).
- A perfect (azimuth 180°) north-south line which connects the axis Phimeanakas-Baphuon with the already existing Bakheng, located outside Angkor Thom at 2.5 Kms (Fig. 4). The latter is built on a hilltop and there is little, if any, doubt that this alignment is intended and therefore governed the longitude positioning of the Phimeanakas-Baphuon complex.

In both cases, it remains to be proved if these alignments are mere survey lines or, as it seems likely actually, there were historical reasons of dynastic and/or explicit reference to tradition which moved the kings to ask for these connections in their state projects.

### 3.2 Sosticial alignments.

At the latitude of Angkor (13° 26' N) the rising azimuths of the sun at the solstices are of 65.5°/114.5°. As mentioned, Paris claimed for the existence of as much as 26 sosticial alignments between monuments. Of these, two refer to the Angkor Wat western entrance and will be discussed separately. I have subjected to an accurate check all the others. A few refer to monuments which are not recognizable; my conclusion on all the others is that they are likely casual. Indeed most alignments refer to secondary monuments and to features which can easily arise from a selection of data. For instance, consider the (very precise indeed!) "sosticial alignment" which connects the north-west corner of the enclosure of Angkor Vat with the north-east corner of the east baray. Clearly such an alignment cannot have any symbolic significance, nor can we imagine an observer who uses these two corners for solar observation. The alignment – in addition – would have no practical utility for a surveyor, and can therefore be definitively discarded. In practice, the only case of a sosticial alignment which survives to this analysis is an alignment connecting the top of Phimeanakas with the centre of Neak Peân, built later. I cannot exclude intentionality, however from Neak Pean – limiting only to main monuments, without considering corners and other features but only tops – one could trace at least some 20 other "alignments" with pre-existing temples.

Regarding Angkor Wat, in their 1976 paper, Robert Stencel, Fred Gifford and Eleanor Moron put in evidence two supposed sosticial alignments already found by Paris. In this way, the notion that Angkor Wat was a sort of calendrical device became widespread. According to this idea, standing at the west gate (the same position from where the equinoctial hierophany can be seen) the sun at summer solstice rises in alignment with the temple located on the Phnom Bok hill some 17.5 Kms to the north-east, while at the winter solstice it rises in alignment with a temple called Kuk Bangro, about 5.5 Kms to the south-east.
Kuk Bangro is a small and damaged ruin which is almost invisible in satellite images, and I actually have doubts that the alignment was ever explicitly verified since Paris found it. The temple is indeed – and with all probabilities has always been - definitively invisible from Angkor Wat; further, its date of construction is not known (and so it could be later than Angkor Wat). All in all, – unless a strict historical connection can be made between the two – the alignment is very likely casual.
Phnom Bok is a hill about 220 meter high, not particularly prominent from Angkor Wat. The temple on its summit was long existing when Angkor Wat was planned, but – although the alignment is indeed verified - to admit intentionality one should assume that the positioning of Angkor Wat as a whole was largely governed by the will of realizing this alignment, a thing for which there is no cultural basis whatsoever.
All in all, the present analysis does not confirm the idea that any of the temples, Angkor Wat included, were used as "calendars in stone".In addition, there is no other kind of evidence showing an interest of the builders of the Angkor monuments for the extreme positions of the sun at the horizon; actually, and naturally, at the latitude and with the climate of Angkor, the interest was rather focussed on the equinoxes and the zenith passages, as associated with the main climatic events, the transitions between the dry and the wet seasons.

**4. The orientation of the two "peripheral" state temples.**

Two state temples were founded out of the Angkor heartland. The first, Koh Ker, was founded by king Jayavarman IV in the mid 10th century AD. The site is characterized by a huge baray and by a 36-meters tall stepped pyramid, which is located in axis with the main temple, the Prasat Thom. The entire project exhibits a peculiar orientation at azimuth 76° (flat horizon) which is shared also by the short sides of the baray. Sometimes topographical reasons - such as the slight south-north slope of the terrain - have been advocated for this orientation (see e.g. Uchida et al. 2014), but it is frankly difficult to believe that the architects of such a huge and complex project might have been influenced by this fact up to rotate the whole design by 14°.
If we search for an astronomical interpretation, an answer is readily found. At the latitude of Koh Ker, azimuth 76° with flat horizon yields a declination of +13° 26'. The latitude of the site is 13° 44' so the main axis is quite precisely oriented to the rising sun on the days of the zenith passages, which of course occur when the sun has a declination equal to the latitude of the place. Therefore, when it was decided to change the place of the state-temple, also a change of orientation was devised, realizing – for the first time – an explicit connection of the axis with the sun rising on the days of the zenith passages (Fig. 6).
Even more clealry, orientation seems to have been a means of expressing a breakthrough when Preah Khan of Kompong Svay was constructed. It is a huge complex: the exterior perimeter of about 5 km per side makes it the largest Khmer enclosure ever built (Mauger 1939). The site was connected to Angkor by a “royal” road rich in stone structures, such as bridges and “rest house” temples (Hendrickson 2010). The building chronology is difficult to establish, since only one dated inscription (to 1010 AD) has been recovered. Accordingly, the site might have been founded in the 11th century by king Suryavarman I. However, important architectural details point to the first half of the $12^{th}$, during which Angkor Wat was also constructed. Curiously, yet other details recall the late $12^{th}$ to early $13^{th}$ century, pointing to king Jayavarman VII, the builder of the Bayon. The religious dedication of the complex is equally difficult to individuate, due to the interplay between Buddhist and Hindu elements.

The causes leading to the construction of such a majestic architecture in such a remote place are still subject to debates, since activities related to iron melting were carried out in this area, so it might have been an administrative centre However, the presence of such an impressive, symbolic monument is difficult to explain, and a complete re-analysis of the archaeological setting together with a new mapping of the area is currently giving new insights into this problem (Hendrickson & Evans 2015).

As far as we are concerned here, there is an aspect which has been noticed by all authors but never explained satisfactorily, namely the anomalous orientation if compared to the Angkor monuments. The complex is indeed clearly rotated to the north of east; Mauger gives 27° 24' north of east, but repeated measures on satellite images rather point to 29° north of east (azimuth 61°). It is this value which will be used here; the horizon is flat or nearly flat.

The hypothesis which has been analysed by some authors (see e.g. Paris 1941) is that the complex might have been orientated to the rising sun at the summer solstice. However the azimuth of the midsummer sun with a flat horizon in this region is 24° 30' north of east and therefore definitively far from the observed one. Of course this does not necessarily mean that the temple was deliberately oriented to another astronomical phenomenon, but topographical reasons are difficult to imagine, and invoking “chance” is equally difficult, also taking into account the strict astronomical pattern which was the rule in Angkor's heartland. Further, a very clear astronomical solution does exist, involving the Moon. As is well known, the plane containing the Earth and the Moon orbit is not the ecliptic, but forms with it an angle of 5° 9'. This has the consequence that the maximal and minimal declinations which the Moon can attain are greater/lesser than those of the sun (which of course equal ± the obliquity of the ecliptic, 23° 30') by such an amount. This leads to the fact that the Moon at the horizon can attain azimuths lesser/greater than those of the sun at the solstices; due to a series of physical reasons however the extreme declinations, also called maximal standstills, are attained only once each every 18.6 years. Of particular interest is the full Moon closest in time to the winter solstice, since it always attains a declination close to the maximal one in the year of the standstill, culminating very high in the sky and remaining in the sky almost the whole night.

Preah Khan of Kompong Svay is definitively oriented to the Moon rising at the maximal northern standstill. Indeed azimuth 61° with a flat horizon at this latitude yields a declination +28°; the true lunar declination at the site is actually slightly greater, 28° 14', because parallax must be taken into account. This value has to be compared with the standstill declination of +28° 39' and again, as is the case of Koh Ker for the zenith passages, the matching is really impressive.[5]
Of course the role of the Moon is quite relevant both in Hinduism – where it is identified with the God Chandra – and in Buddhism, since festivals and recurrences associated with Buddha's life are timed by the full Moon. The choice of orienting to the extrema of the Moon might thus have stemmed from this and/or from other specific messages the builder wanted to associate with the temple. Of course, further archaeological/historical researches are necessary to clarify this point. There is, however, a second issue related to astronomy which is worth discussing about Preah Khan of Kompong Svay. It is in fact known that the temple is located on the same latitude of Angkor Vat. The accuracy of this coincidence is astonishing: the centre of Angkor Vat is within 1' at the same latitude (13° 24' ) of the entrance to the Preah Khan inner enclosure. The linear distance between Angkor Vat and Preah Khan of Kompong Svay is of about 100 Kms, and therefore no visual connection is conceivable, and it is definitively possible to think that the connection occurs by chance. Indeed, to proceed along a "straight" line between non-intervisible positions is a thing which generally has no sense at all since of course there are no straight lines on a curved surface (as pseudo-archaeologists in search of "leylines" usually forget). However, the special case of sites placed at the same latitude *does* make sense, because – although the parallel circle of course is *not* the shortest path between two points at the same latitude - the parallel in itself is (in principle) easy to determine using astronomy, and the Khmer architects certainly had the necessary skills. It is indeed possible to establish latitude by measuring the height of the celestial pole or the height of the sun at midday in fixed days, this second method being clealry favoured in the present case. Ancient people interested in the zenith passages of the sun actually appear to have been also interested in developing precise measures of latitude, as shown, for instance, by the archaeological site of Altavista in Central Mexico. However, even admitting intentionality, obtaining the accuracy exhibited by the Angkor- Preah Khan alignment must have been quite a daunting task, which clealry demands for a sound historical explanation and cries out for a symbolic, rather than functional, interpretation for the site in connection with Angkor Wat.

## 6. Discussion

The Angkor temples are masterpieces, built by knowledgeable architects who planned and erected them taking into account a complex symbolical framework connected with the explicit representation of the ruler's power and of his relationships with the Gods. The results of the present paper show that this framework included the sky: it was a worldview, where the cycle of the sun and that of the dry and the wet seasons were tightly connected. Interest was mainly focussed on the equinoxes and on the zenith passages, as both phenomena were implemented in the temple's architectural features and projects.

[5] The variation of the obliquity of the ecliptic since 1100 AD is negligible, being about 6'.

In a few, but relevant cases also the temples by themselves were not isolated units but were ideally linked with pre-existing monuments, constructing a series of visually recognizable, dynastic lines which are particularly evident in the case of the Mebons, the island temples. It is thus the hope of the author that the present research can contribute to clarify historical aspects of the Khmer architecture and king's succession. On the opposite side, the same results show that claims about the existence of scores of inter-connecting, almost esoteric lines between the Angkor monuments must be taken with the utmost care, if not definitively refused.

**Acknowledgements**

The author gratefully acknowledges Dr. Damian Evans and the *Greater Angkor Project* for their kind permission to double-check the author's data with the GIS database they are developing.

| MAIN MONUMENTS OF THE ANGKOR AREA | Orientation | Notes | King/date |
|---|---|---|---|
| **Preah Ko** | 90 180 | | Indravarman I<br>880 |
| **Bakong** | 89 179 | State temple | Indravarman I<br>881 |
| **Baray Indratataka** | 90 179 | | Indravarman I |
| **Lolei** | 90 180 | Island temple | Yasovarman I<br>893 |
| **Phnom Bakheng** | 89,5 179 | State temple | Yasovarman I<br>907 |
| **Baray East** | 88 178 | | Yasovarman I |
| **Phnom Krom** | 90 180 | | Yasovarman I |
| **Phnom Bok** | 90 179 | | Yasovarman I |
| **Prasat Bei** | 90 180 | | Yasovarman I |
| **Baksei Chamkrong** | 90 180 | | Harshavarman |
| **Prasat Kravan** | 90 179 | | Harshavarman I<br>921 |
| **East Mebon** | 86 177 | Island temple | Rajendravarman II<br>953 |
| **Pre Rup** | 88,5 179 | | Rajendravarman II<br>961 |
| **Bat Chum** | 90 180 | | Rajendravarman II |
| **Banteay Srey** | 90 180 | Private construction | Rajendravarman Ii<br>967 |
| **Phimeanakas** | 89 179 | Date and king unsure | Rajendravarman II |
| **Ta Keo** | 89 180 | | Jayavarman V<br>1000 |
| **Chau Srei Vibol (Wat Trak)** | 89,5 179,5 | Date and king unsure. Difficult to measure | Suryavarman I<br>1000-1050 |
| **Baray West** | 90 180 | | Suryavarman I |
| **Baphuon** | 89 179 | | Udayadityavarman II<br>1050-1066 |
| **West Mebon** | 90 180 | Island temple | Udayadityavarman II<br>1050-1066 |
| **Angkor Wat** | 270,5 180 | State temple, faces west | Suryavarman II<br>1113-1145 |
| **Banteay Samre** | 85 175 | | Suryavarman II<br>1113-1145 |
| **Wat Athvear** | 270 180 | Faces west | Suryavarman II<br>1113-1145 |
| **Thommanon** | Non meas. | | Suryavarman II<br>1113-1145 |
| **Chau Say Tevoda** | Non meas. | | Suryavarman II<br>1113-1145 |
| **Beng Melea** | 89,5 179 | | Suryavarman II<br>1113-1145 |
| **Ta Prohm** | 87,5 177,5 | | Jayavarman VII<br>1181-1218 |
| **Preah Khan** | 89 179 | | Jayavarman VII<br>1113-1145 |
| **Baray Preah Khan** | 89 178 | | Jayavarman VII<br>1113-1145 |
| **Neak Pean** | 88.5 178.5 | Island temple | Jayavarman VII<br>1113-1145 |
| **Bayon** | 89,5 179 | | Jayavarman VII<br>1113-1145 |

| | | |
|---|---|---|
| **Banteay Kdei** | 85 175 | Jayavarman VII<br>1113-1145 |
| **Baray Banteay Kdei** | 86 176 | Jayavarman VII<br>1113-1145 |
| **Ta Som** | 88  178 | Jayavarman VII<br>1113-1145 |
| **Krol Ko** | 87 177 | Jayavarman VII<br>1113-1145 |
| **Banteay Prei** | 88 178 | Jayavarman VII<br>1113-1145 |

**STATE TEMPLES**
**OUT OF ANGKOR HEARTLAND**

| | | |
|---|---|---|
| **Preah Kahn of Compong Svay**<br>**Baray** | 60 150<br>61 151 | |
| **Kok Ker**<br>**Baray** | 76 164<br>76 165 | |
| **Banteay chhmar**<br>**Baray** | 88 178<br>88 178 | Jayavarman VII<br>1113-1145 |

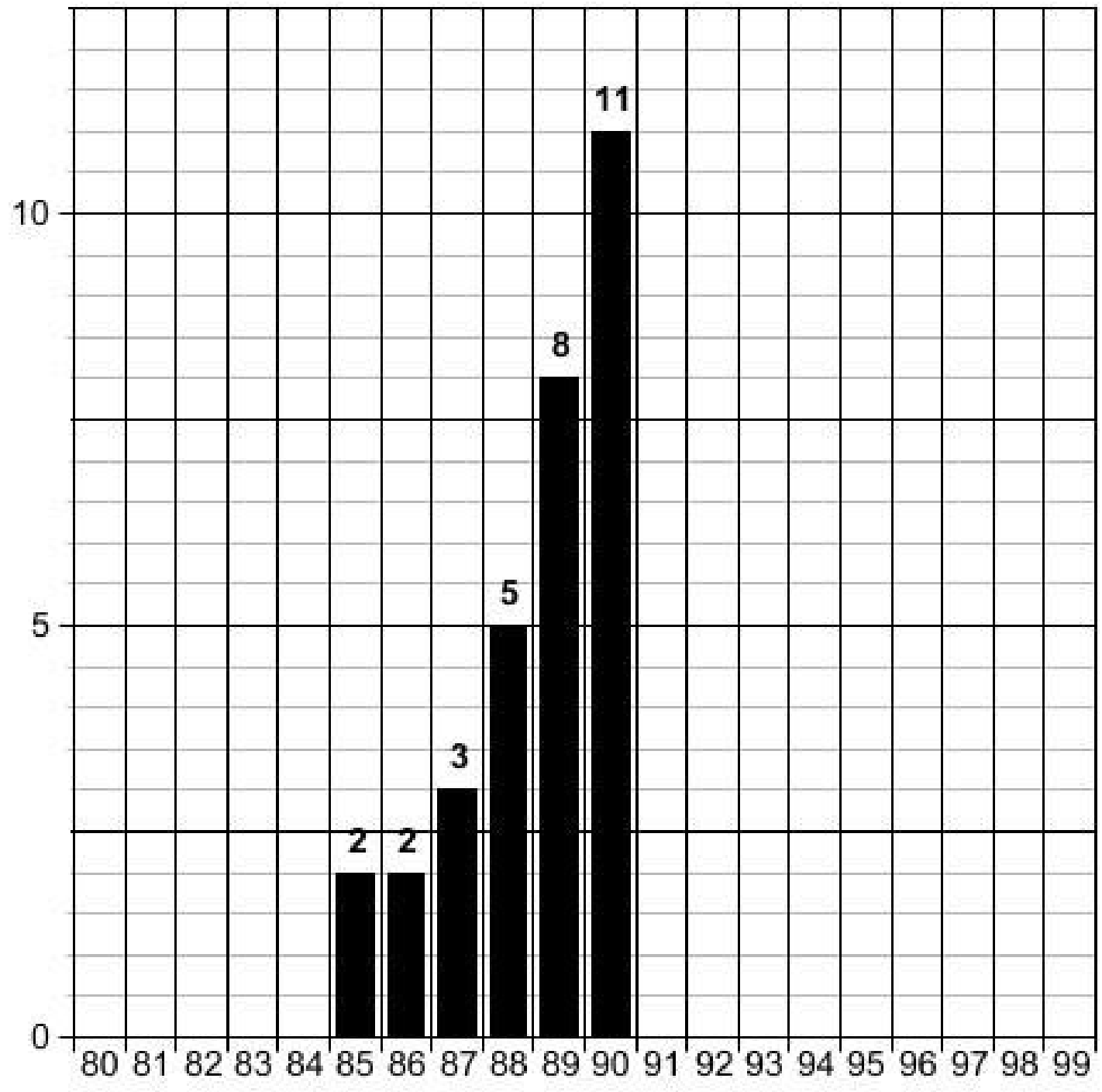

**Fig. 1. Orientation histogram (azimuth in degrees vs. number of monuments) of 31 monuments of the Angkor Heartland.**

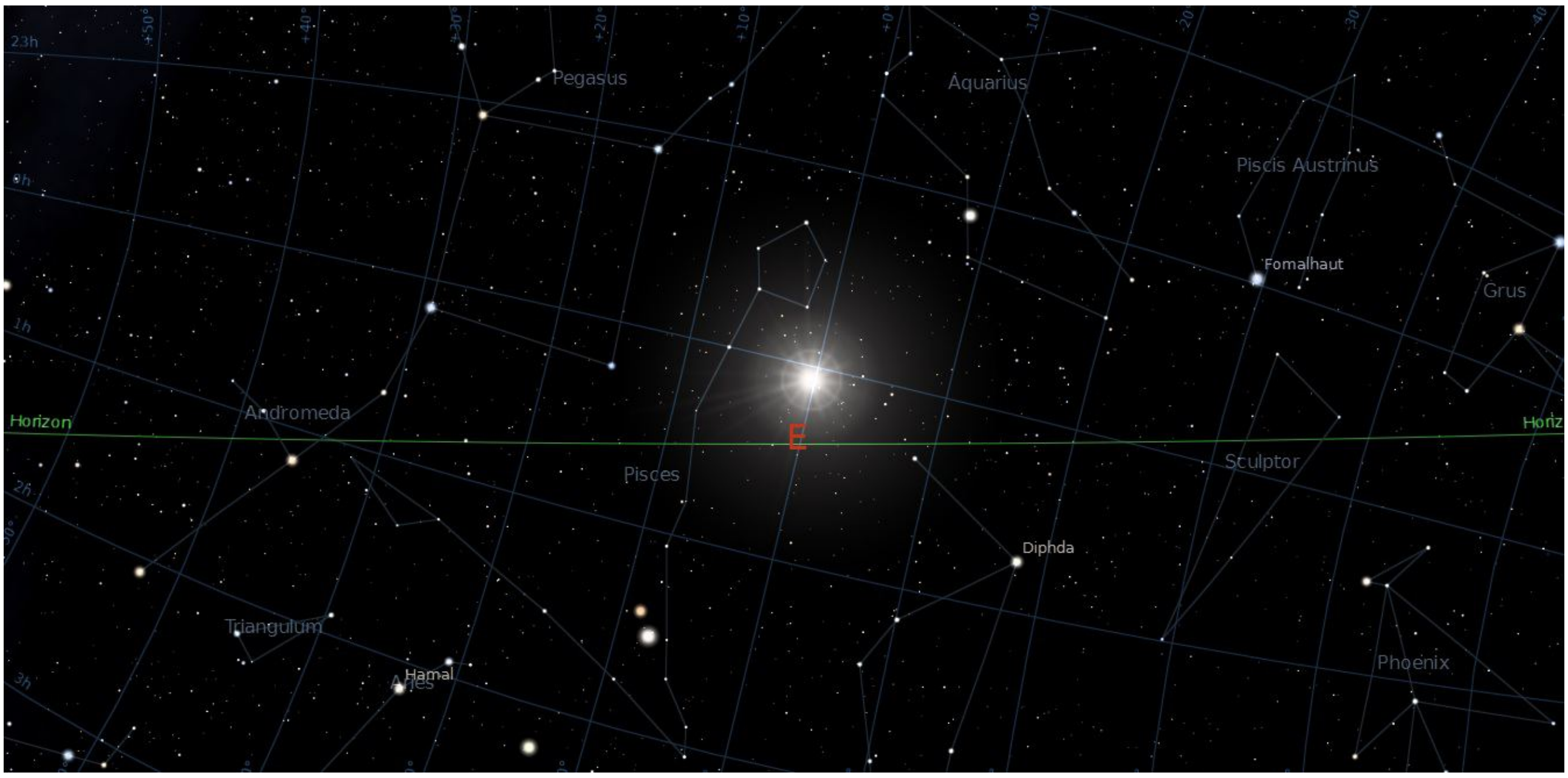

**Fig. 2. Stellarium simulation of the sun rising at the Spring equinox at Angkor. The coordinates of the sun are: azimuth 90° 40', height 5°.**

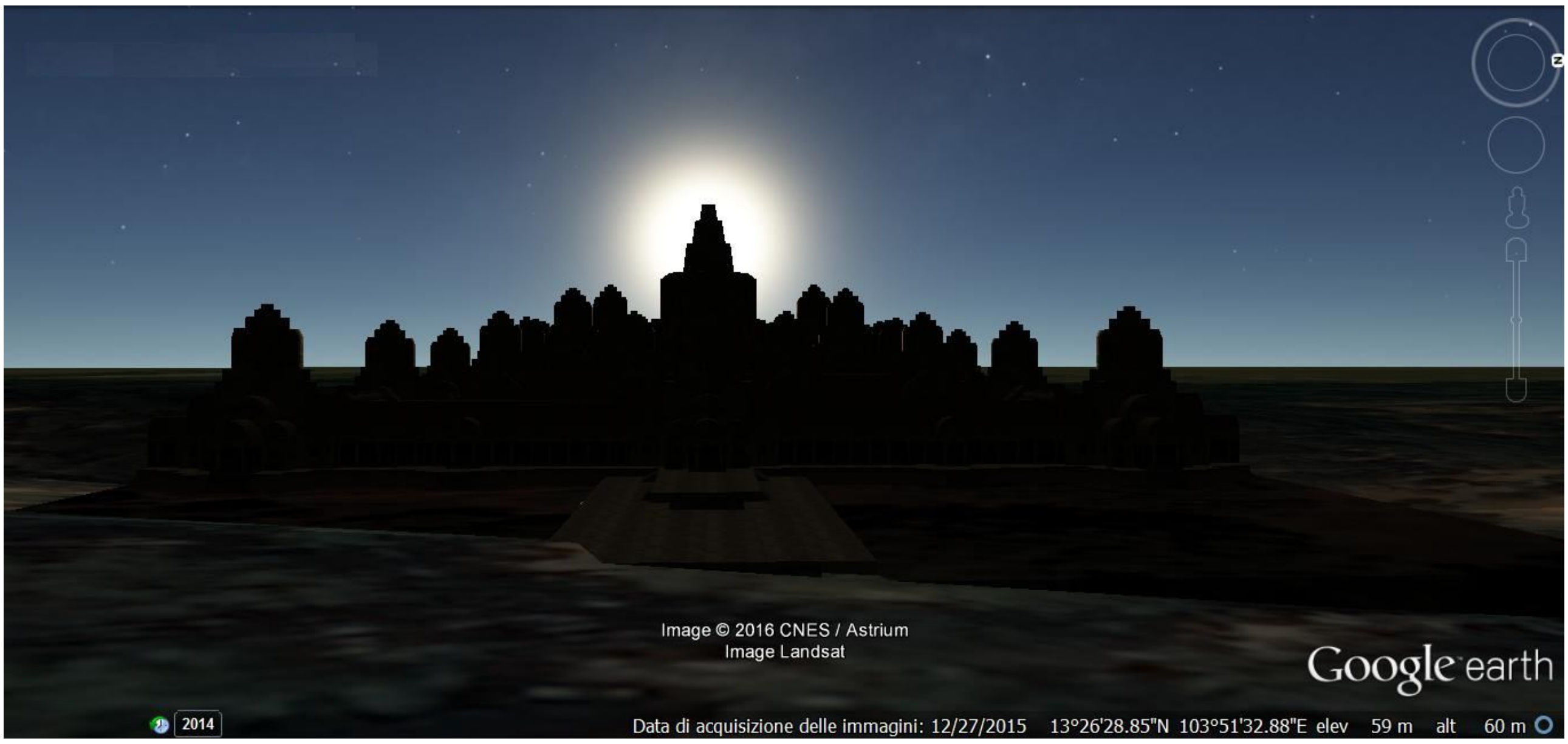

**Fig. 3. Google Earth simulation of the setting of the sun behind the Bayon at spring equinox.**

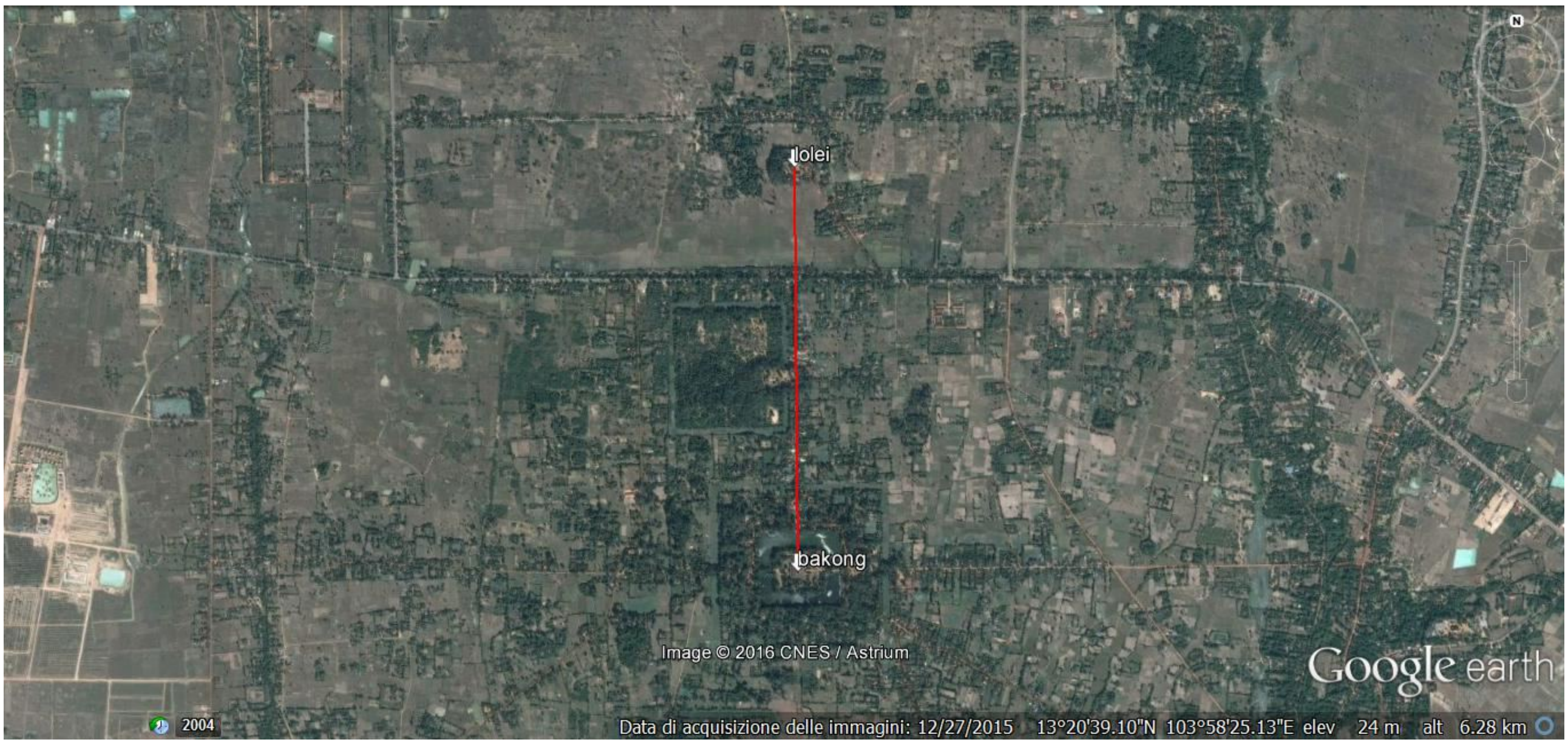

**Fig. 4. The meridian line connecting Lolei with the Bakong (Image courtesy Google Earth, drawings by the author)**

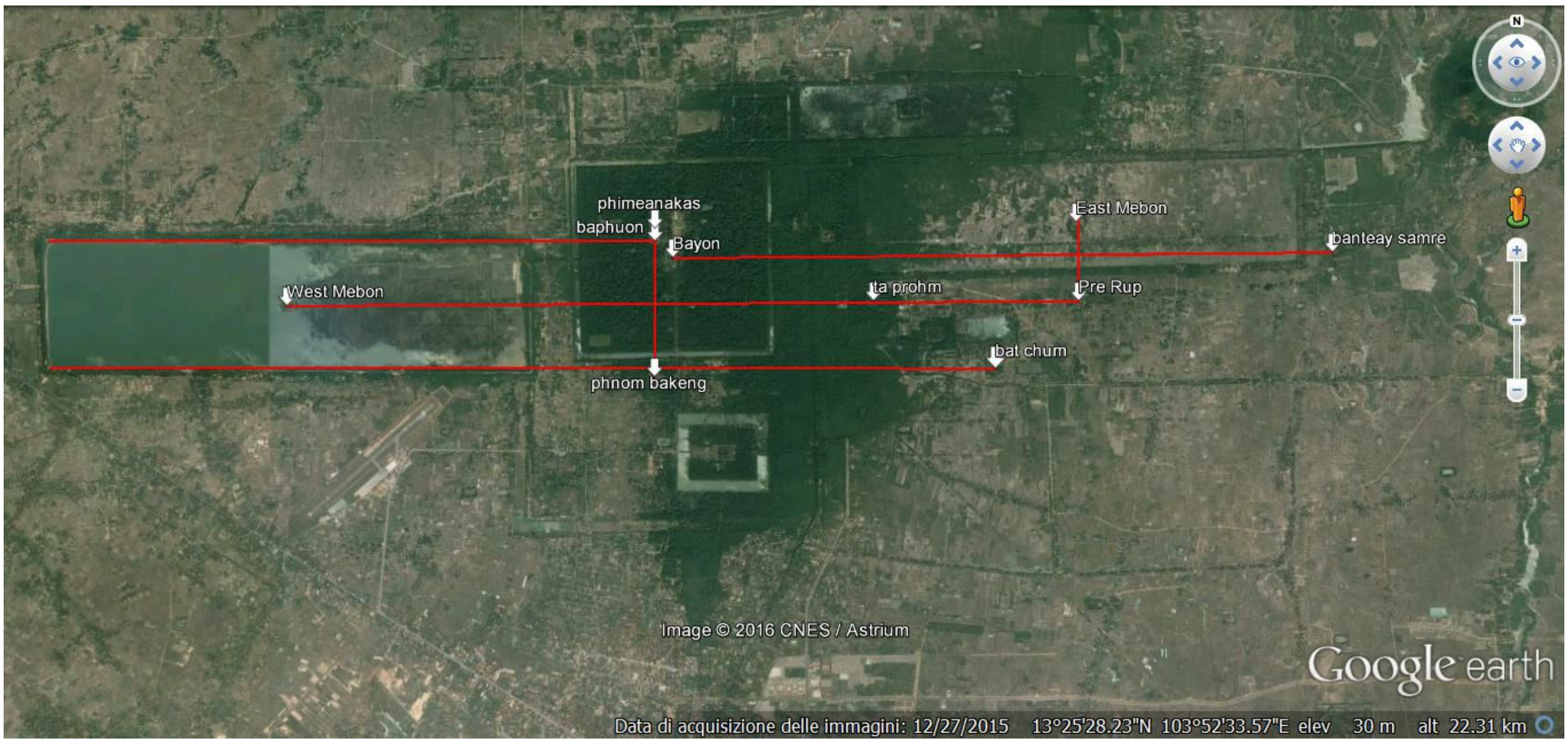

**Fig. 5. Cardinally oriented aligments between main monuments at Angkor (Image courtesy Google Earth, drawings by the author)**

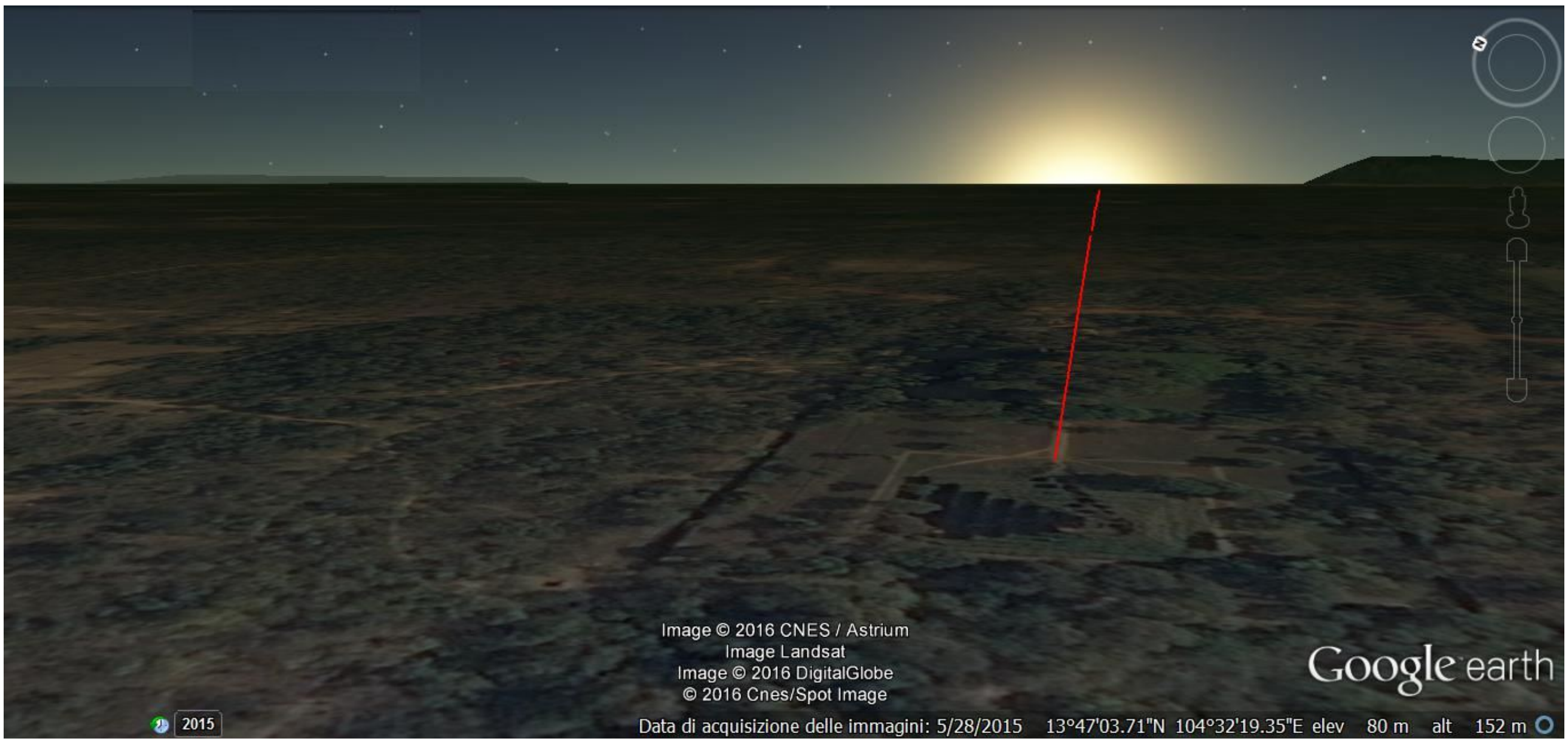

**Fig. 6. Google Earth simulation of the sun rising in alignment with Kok Ker on the day of the first zenith passage (Image courtesy Google Earth.**